\documentclass[twocolumn,showpacs,preprintnumbers,amsmath,amssymb]{revtex4}
\usepackage{graphicx}
\usepackage{bm}
\usepackage{epsf}
\usepackage{xcolor}
\usepackage{epstopdf}
\begin{document}
\title{Melting of thin silicon films: a molecular dynamics study with two machine learning potentials}

\author{Yu. D. Fomin \footnote{Corresponding author: fomin314@mail.ru}}
\affiliation{Vereshchagin Institute of High Pressure Physics,
Russian Academy of Sciences, Kaluzhskoe shosse, 14, Troitsk,
Moscow, 108840, Russia }

\author{E. N. Tsiok}
\affiliation{Vereshchagin Institute of High Pressure Physics,
Russian Academy of Sciences, Kaluzhskoe shosse, 14, Troitsk,
Moscow, 108840, Russia }

\author{V. N. Ryzhov }
\affiliation{Vereshchagin Institute of High Pressure Physics,
Russian Academy of Sciences, Kaluzhskoe shosse, 14, Troitsk,
Moscow, 108840, Russia }
\date{\today}

\begin{abstract}


Thermal stability of silicene and thin silicon films is studied by molecular dynamics using two machine-learning potentials, SNAP and GAP.
For SNAP potential, systems ranging from a single silicene layer to films of 36 layers are considered. Silicene is found to lose its structure at
500 K. The decomposition temperature increases with film thikness and reaches saturation at about 28 layers, corresponding to the bulk melting point of 
the SNAP model (1380 K). Thin films up to 8 layers exibit two-phase coexistence upon decomposition, while thicker films undergo surface melting 
followed by complete collapse into the liquid state. The GAP potential, although more accurate for bulk silicon, fails to describe the gas phase:
silicene modelled with GAP decomposes into a set of small clusters. The results are compared with earlier data for the Stillinger-Weber potential.

\end{abstract}


\maketitle


\section{Introduction}

Silicon is the base of modern electronics. It makes this material of great technological importance. In particular,
silicon structures with low dimensionality, such as silicene, are interesting as promising two-dimensional materials.

Silicene was predicted in a theoretical work \cite{si-1} in 1994 and synthesized in 2012 \cite{si-syn}. Unlike graphene,
silicene is not a perfectly flat material, but a corrugated layer, i.e., some of the atoms are above the symmetry plane
and others are below. It is expected that silicene can be employed in different technological applications, including
electronics, which is a traditional field for silicon \cite{si-electronics}, but also promising technologies
of green energy, such as hydrogen storage \cite{si-h}, lithium-ion batteries \cite{si-li-ion}, etc.


Although the properties of silicene were widely studied both experimentally (see the review articles \cite{rev-1,rev-2} and
references therein) and in simulations (see, for instance, \cite{dft-1,dft-2,dft-3,dft-4,dft-5,dft-6} for
density functional theory studies of silicene), there are still a lot of opened questions. It is of particular interest to study
the thermal stability of silicene. At the same time this is a difficult problem both for experiments and simulations.
The melting of silicene and thin silicon film is challenging for DFT calculations since it requires rather large
samples (thousands of atoms) at rather high temperature. For this reason it looks appropriate to employ
the methods of molecular simulation with effective potentials, which allow to calculate even millions of atoms.
However, these effective potential can be not accurate enough to reproduce the behavior of a real system. It results
in a great spread of data for the thermal stability of silicene available in the literature. For instance,
melting of silicene modelled with the Tersoff potenital \cite{tersoff} was reported in Ref. \cite{melting-1}. Two sets
of parameters of the potential were used: the original one \cite{tersoff} and the one developed for the
void-nucleated melting of silicon \cite{tersoff-a}. The melting temperature was estimated to be $T_m=3600$ K
in the former case and $T_m=1750$ K in the latter. This example clearly show that the melting point of silicene
is extremely sensitive to the model used.

In our recent papers we employed a well recognized Stillinger-Weber model \cite{sw} to simulate the thermal stability
of silicene and thin silicon films of different width \cite{we1,we2}. It was found that the temperature of thermal
decomposition of a silicon film increases with increasing of the film width. However, when the number of layers
is larger than 12 the decomposition temperature comes to a saturated value consistent with the melting temperature of
bulk silicon: $T_=1725$ K (the experimental value is $T_m^{exp}=1683$ K \cite{tm}).

As it was discussed above, the estimation of thermal stability of silicene evaluated withing the framework
of empirical force fields strongly depends on the employed model. In this paper we continue the study
of thermal decomposition of silicene and thin silicon films by employing two machine-learning potentials:
SNAP \cite{snap-silicon} and GAP \cite{gap-silicon} models. It is expected that machine-learning models
should better reproduce the interaction of the atoms which should lead to more accurate data.

In our recent study \cite{melt-sn} a comparison of melting lines of SNAP and GAP models was given (the
melting line of GAP model was taken from Ref. \cite{gap-silicon}). Both models give correct qualitative
behavior of the melting line of the diamond phase of silicon (melting line of silicon
is a straight line with negative slope), but both lines are systematically lower than the experimental one.
The melting temperature of silicon at ambient pressure obtained with SNAP model is $T_m=1380$ K which
is substantially below the experimental value. Based on the fact that these models give correct qualitative
behavior, we perform simulations of thermal decomposition of silicene and thin silicon films and compare
the qualitative results with our previous calculations with empirical SW model.

\section{System and Methods}

In the present work we investigate crystal structure collapse of a silicon sample of different thicknesses by means of molecular dynamics simulation.
Two machine-learning potentials are used: SNAP \cite{snap-silicon} and GAP \cite{gap-silicon}. A film of a given thickness is surrounded by vacuum.
Periodic boundary conditions are employed in all directions. The size of the system in z direction, where vacuum is added, is $543$ $\AA$, which is
approximately one hundred times the lattice constant of silicon. Such vacuum gap is sufficient to prevent interaction of the film with its periodic images.
The unit cell parameters of silicene are $a=b=3.861$ $\AA$, and $\alpha=\beta=90^o$ and $\gamma=120^o$, therefore, the box is triclinic.
A rectangular box with equal dimensions in X and Y directions was employed for the thin films with diamond structure. The number of
atoms used in simulations with SNAP model in all considered systems is given in the Table I.

\begin{table}
\begin{tabular}{ l c r }
  System & Number of particles \\
\hline
  silicene & 3200  \\
\hline
  4 & 2048  \\
\hline
  8 layers  & 4096 \\
\hline
 12 layers & 6144 \\
\hline
 16 layers & 8192 \\
\hline
 20 layers & 10240 \\
\hline
 24 layers & 12288 \\
\hline
 28 layers & 14336 \\
 \hline
 32 layers & 18432 \\
 \hline
 36 layers  & 20480 \\
\hline
\end{tabular}
\caption{Number of atoms in systems with different number of silicon layers.}
\end{table}

In the case of GAP model only a system of silicene with 1800 atoms was used. The reasons for this will be described below.

In all cases we start from ideal crystal structure and simulate the system in canonical ensemble (constant number of particles N, volume V and temperature T) at given temperature.
The set of temperatures depends on the system. The goal of this work is to find out at which temperature the crystalline structure collapses. The time step
is 1 fs and all simulations are conducted up to 100 ps.

All simulations were performed with lammps simulation package \cite{lammps}.

\section{Results and Discussion}

\subsection{SNAP model}

\begin{figure}

\includegraphics[width=8cm]{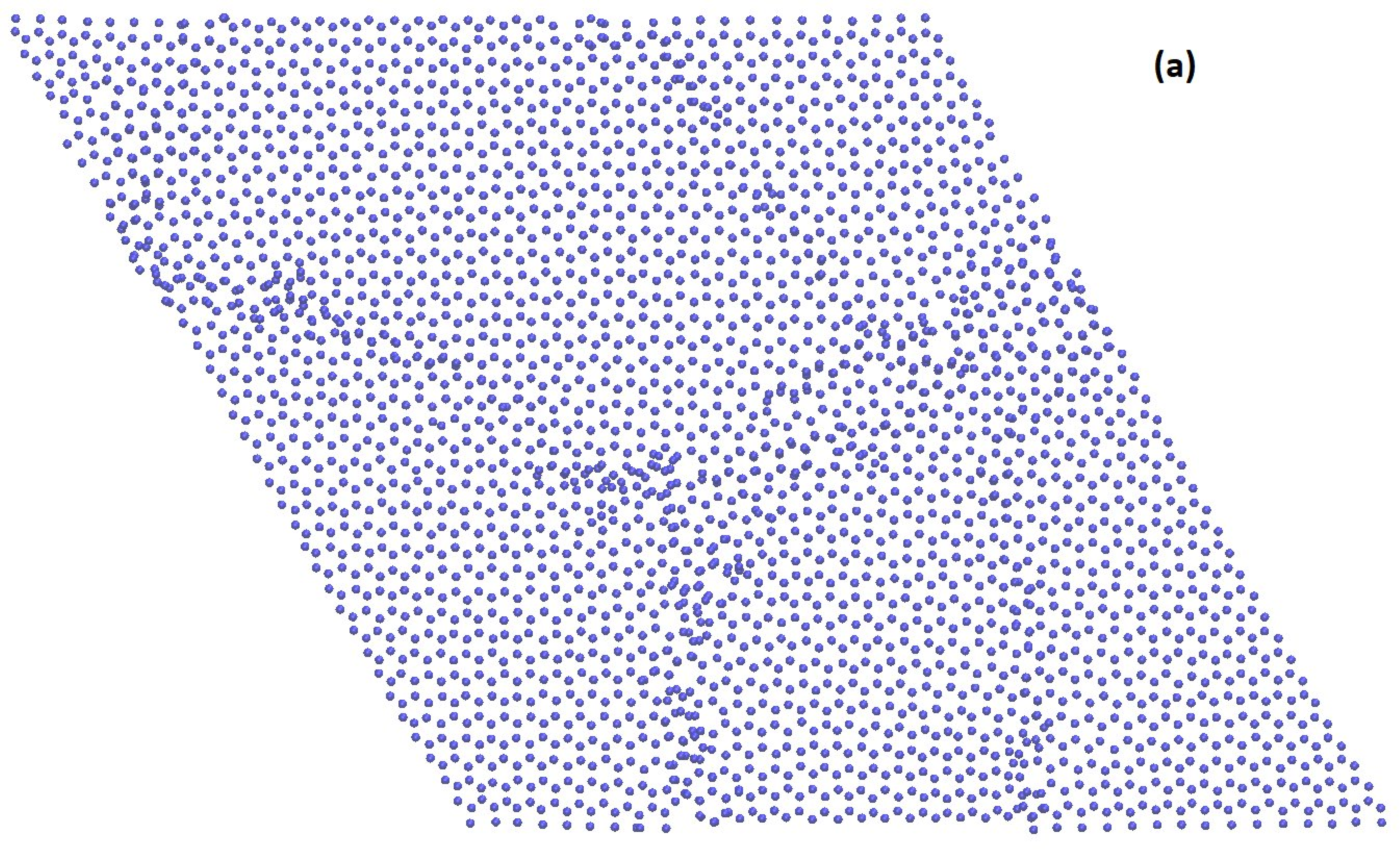}%

\includegraphics[width=8cm]{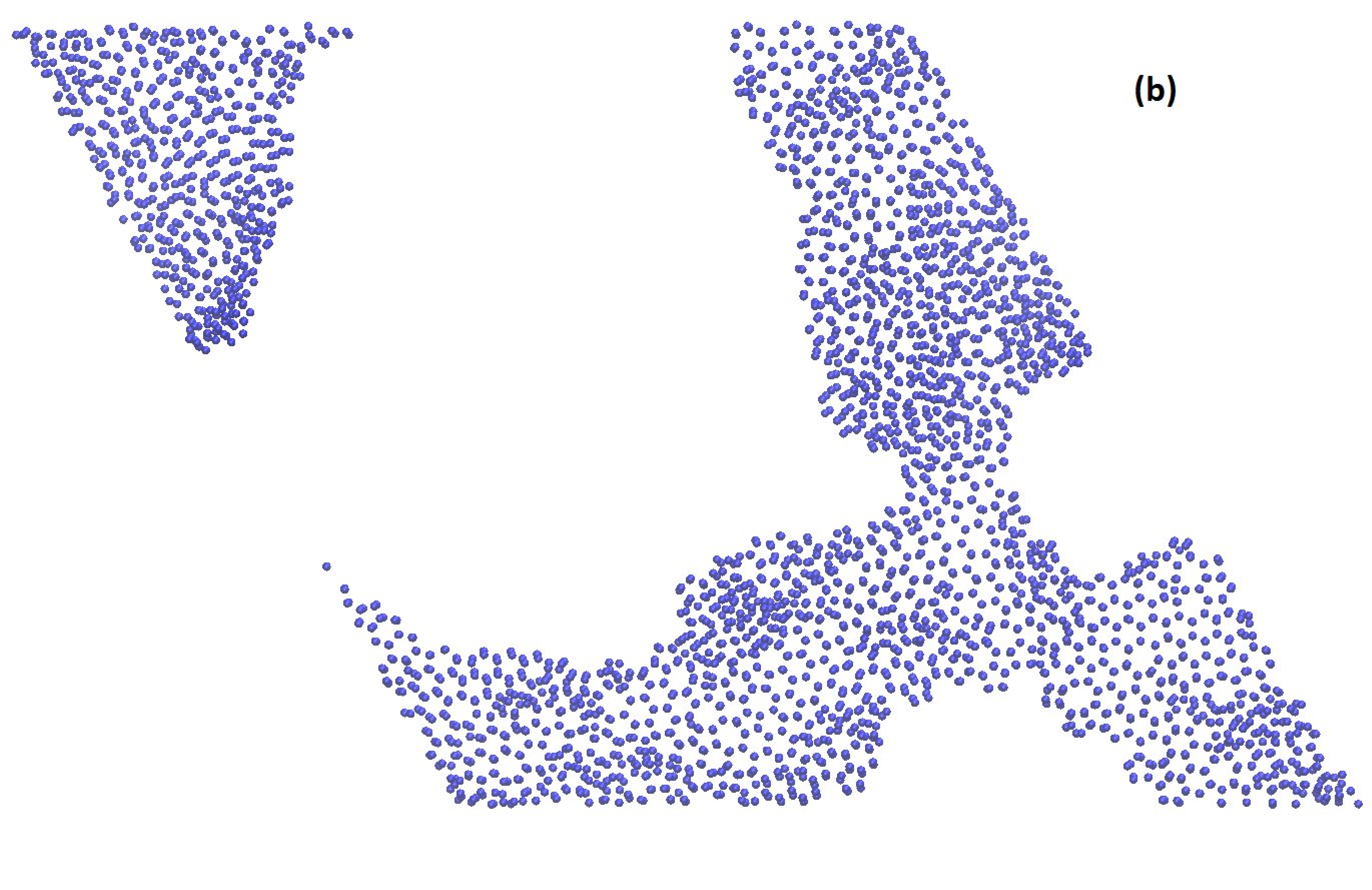}%

\caption{\label{sil-sn} Snapshots of silicene at (a) $T=475$ K and (b) $T=500$ K.}
\end{figure}

First of all we consider the case of silicene. Figure \ref{sil-sn} shows snapshots of the system
at two temperature values: $T=475$ and $500$ K. One can see that at $T=475$ K the silicene
layer is preserved, although it experiences strong corrugation due to thermal motion, but the layer
is broken at $T=500$ K. At the same time some crystalline particles are still observed in the system. One can
conclude that this point corresponds to a two-phase region crystal-gas.

\begin{figure}

\includegraphics[width=8cm]{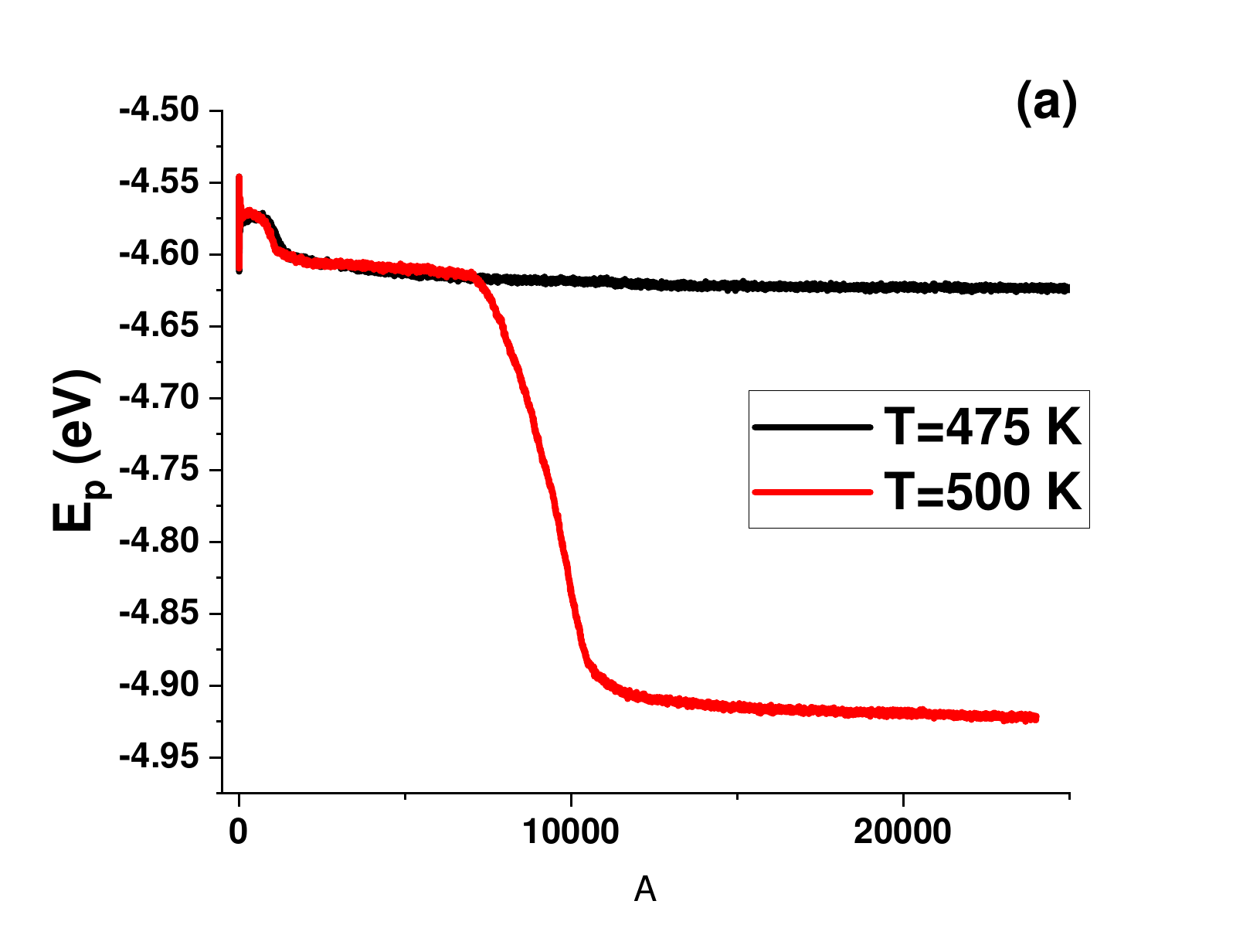}%

\caption{\label{esil} The time dependence of potential energy per particle of silicene $T=475$ K and $T=500$ K.}
\end{figure}

Figure \ref{esil} shows the time dependence of the potential energy per particle
at the same temperatures. Although these temperatures are very close, it is seen that the behavior of the potential
energy is different. While in the former case the potential energy smoothly goes to some limiting value, a rapid drop of the
energy is observed in the latter. This energy drop should be related to a phase transition in the system.

So, the silicene film appears to be stable up to $T=500$ K.

\begin{figure}

\includegraphics[width=8cm]{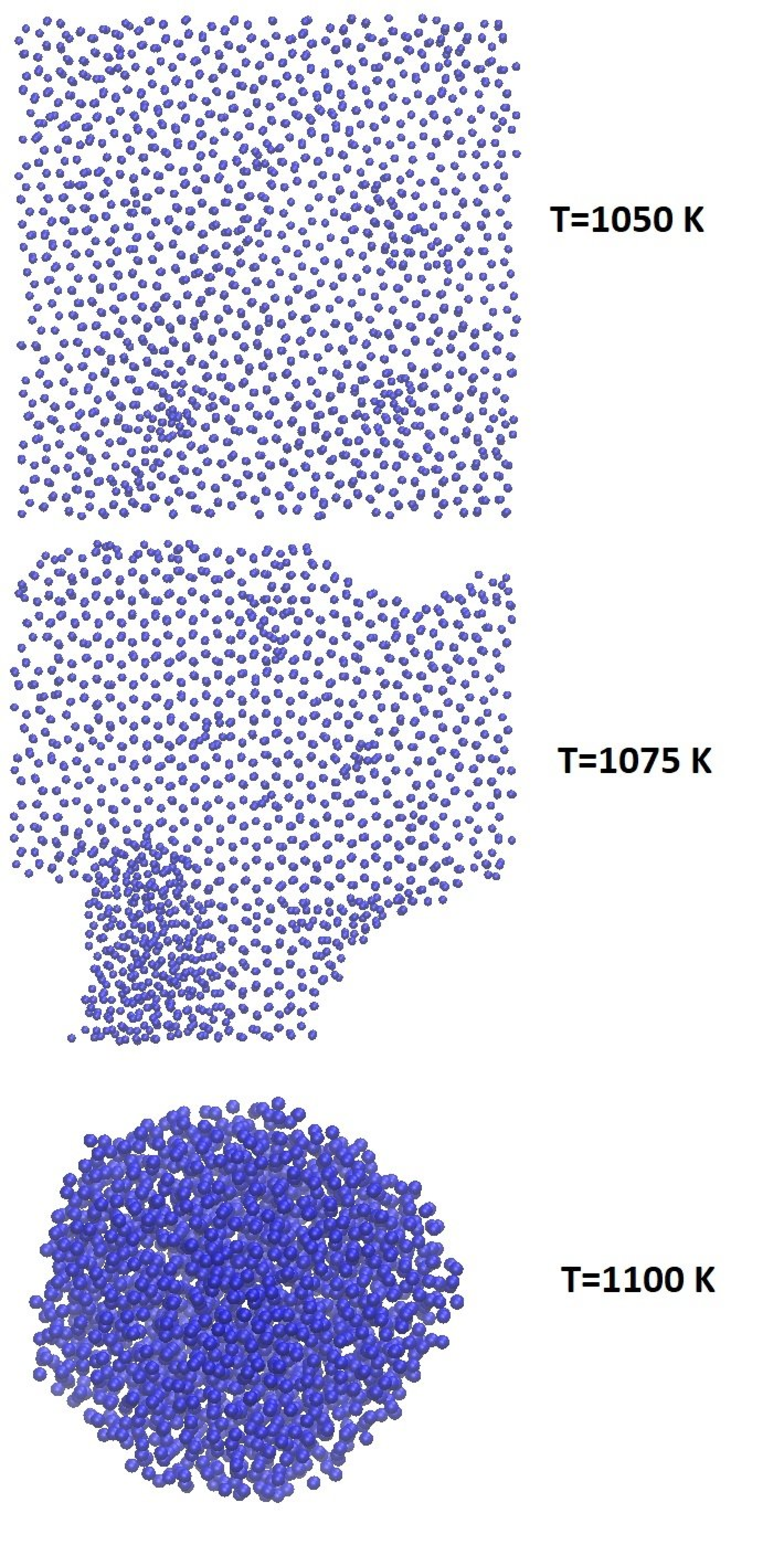}%

\caption{\label{si1} Snapshots of the four layers system at three different temperatures.}
\end{figure}

Let us consider a sample which consists of four layers of silicon. Several snapshots of this system are shown in Fig. \ref{si1}.
When simulated at finite temperature, it recrystallizes in
kinds of graphene-like layers. These layers are stable up to $T=1050$ K. At $T=1075$ K the sample demonstrates some kinds of voids
which means that solid-gas coexistence takes place. Finally, at $T=1100$ K the system collapses into a ball, which corresponds
to the liquid-gas two-phase region. From these figures we conclude that the solid film is stable up to $T=1050$ K.

\begin{figure}

\includegraphics[width=8cm]{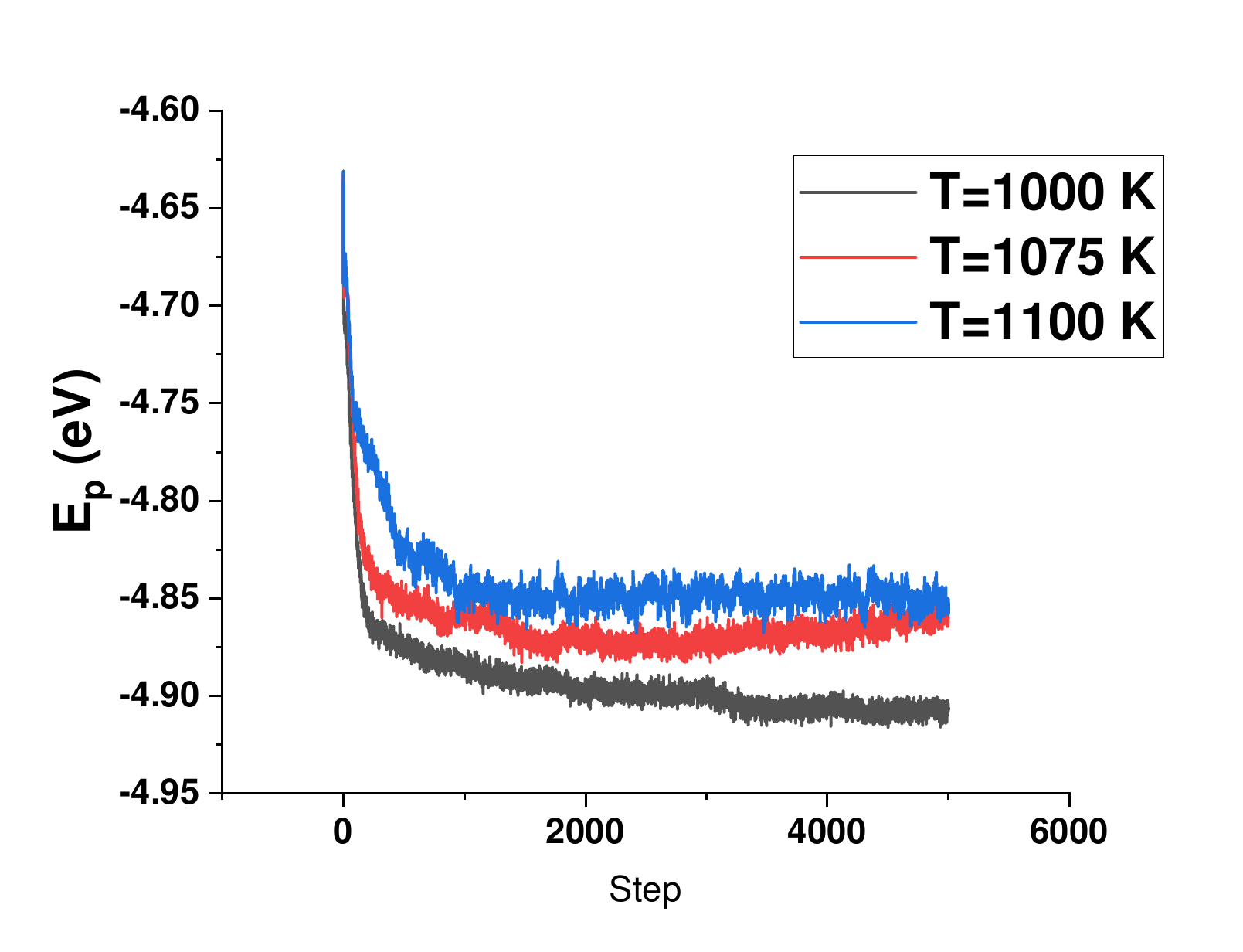}%

\caption{\label{esi1} The time dependence of potential energy per particle of four layer sample at three temperatures.}
\end{figure}

The evolution of the energy of the system is shown in Fig. \ref{esi1}.
It is seen that at $T=1000$ K the energy smoothly relaxes to an equilibrium value, while at higher temperatures some shoulders
are observed in the time dependence of the energy. This kind of behavior is characteristic for the phase transitions in the
system.

\begin{figure}

\includegraphics[width=8cm]{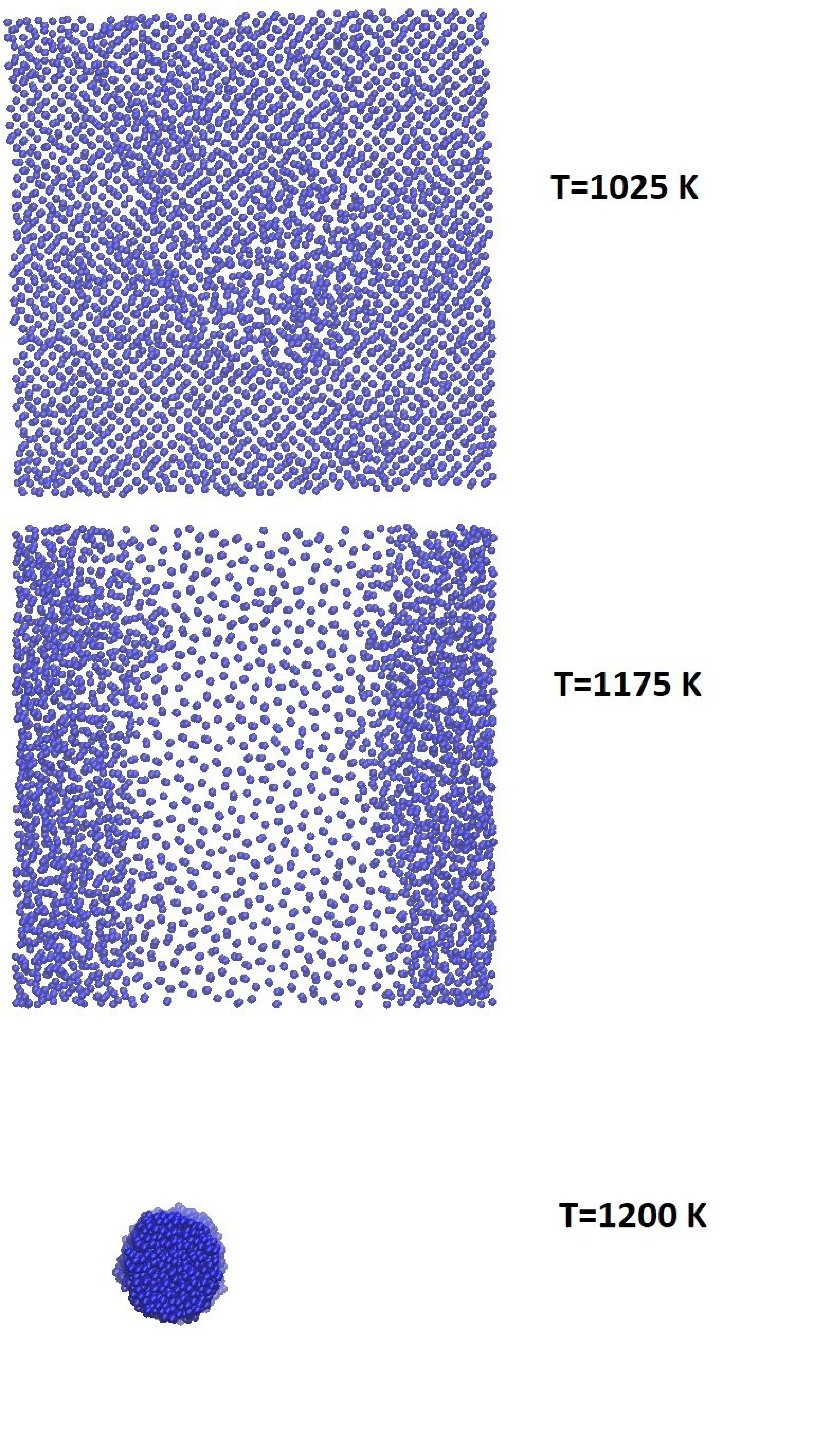}%

\caption{\label{si2} Snapshots of the eight layers system at three different temperatures.}
\end{figure}

The behavior of eight layer system is analogous. At low temperature the structure of the system preserves. At
$T=1050$ K the structure firstly melts, but then partially crystallizes into the graphene-like layers. When the
temperature reaches $1200$ K the structure fails completely. This is seen in the snaphots of the system
in Fig. \ref{si2}. In the case of $T=1200$ K all particles of the system collapse into a cylinder.
This structure is characteristic for a liquid-gas two-phase region in finite size systems with periodic
boundary conditions \cite{prest}.

\begin{figure}

\includegraphics[width=8cm]{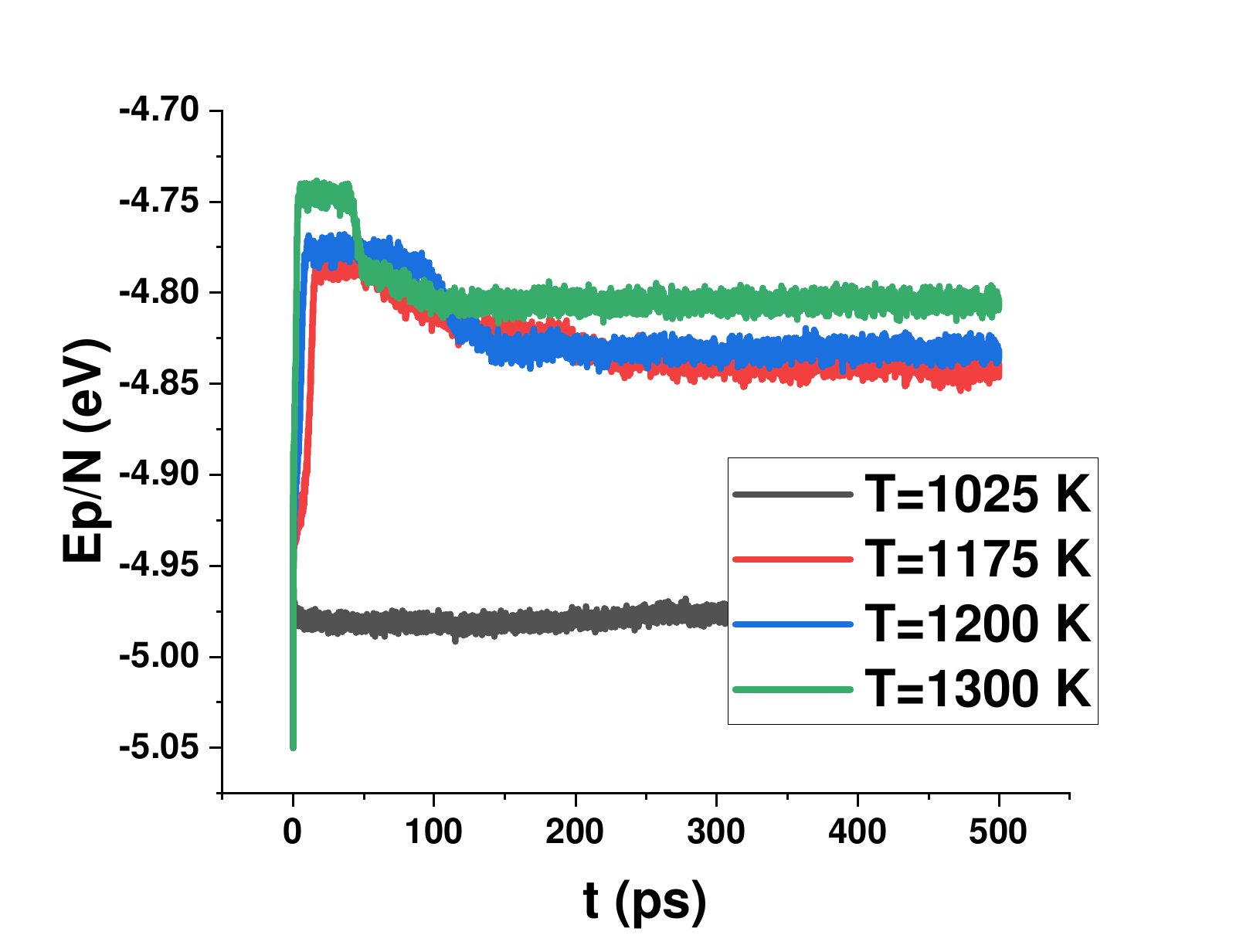}%

\caption{\label{esi2} The time dependence of potential energy per particle of eight layer sample at four temperatures.}
\end{figure}

This can be seen in the energy vs. time curves for this system, which are shown in Fig. \ref{esi2}.
Once again, drops in the potential energy corresponding to phase transitions take place at higher temperatures.


The systems with 16 and more layers of silicon do not demonstrate gas or liquid-gas coexistence. In these systems surface melting of the crystal
takes place. As the temperature becomes higher the surface melting penetrates deeper in the crystal. Finally, when the width
of the crystalline part becomes too thin the crystal rapidly melts. At the same time the system preserves to be condensed, i.e.,
after the crystal structure collapse all particles of the system are located in the liquid phase.

\begin{figure}

\includegraphics[width=8cm]{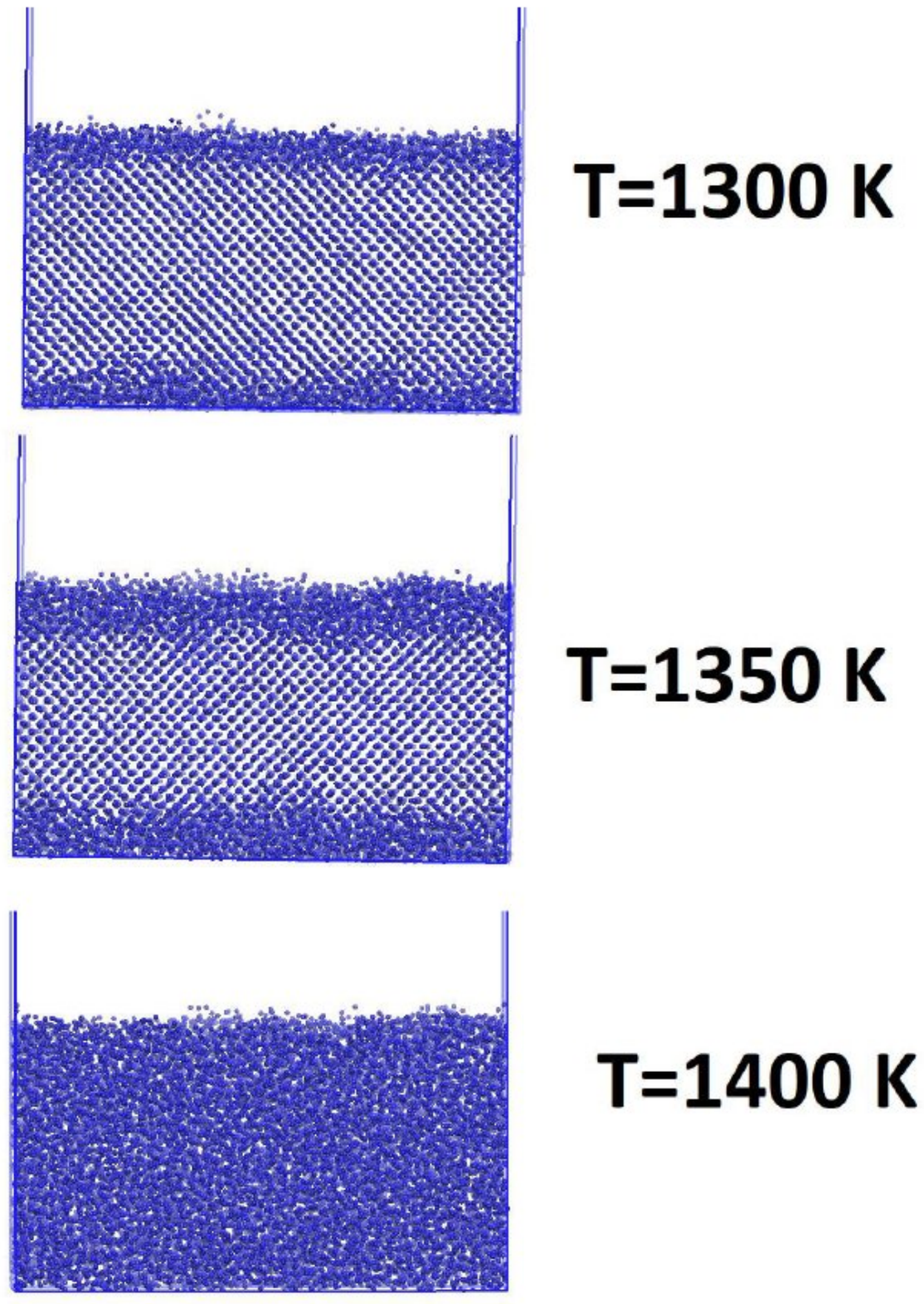}%

\caption{\label{si9} Snapshots of the 36 layers system at three different temperatures.}
\end{figure}

As an example of such behavior we show snapshots of the system with 36 layers of silicon. Snapshots of the system at three
different temperatures are shown in Fig. \ref{si9}. It is seen that the surfaces of the crystal become disordered.
It is also seen that the depth of the disordered surface is larger at $T=1350$ K comparing to $T=1300$ K. However,
the effect is rather small. At the same time at $T=1400$ K the melting appears at the surface of the crystal, but it
proliferates until the melting fronts from the two surfaces meet.

\begin{figure}

\includegraphics[width=8cm]{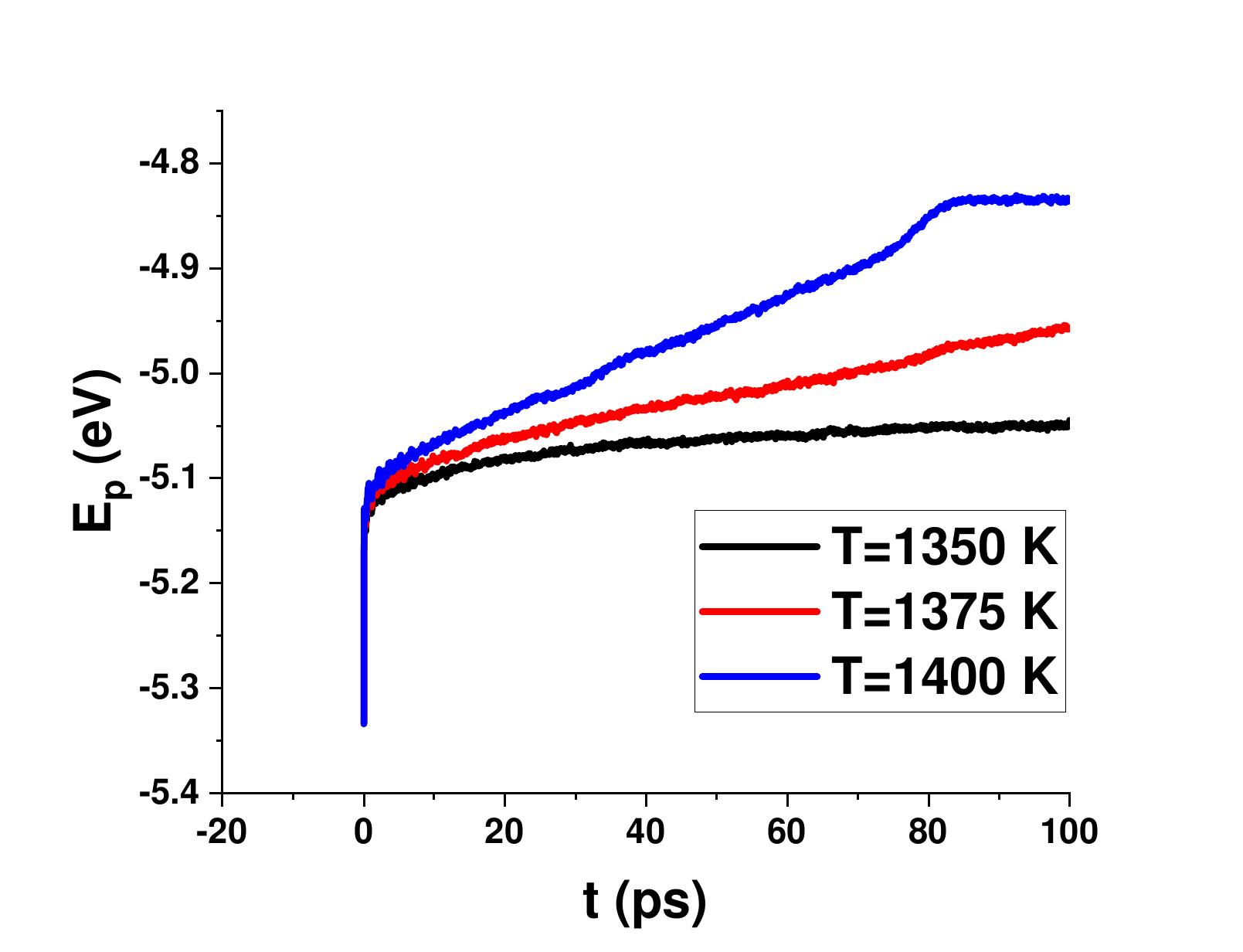}%

\caption{\label{esi9} The time dependence of potential energy per particle of 36 layer sample at three temperatures.}
\end{figure}

The time dependence of the potential energy per particle is given in Fig. \ref{esi9}. No sudden change of the potential energy
is observed in this case. In the case of $T=1350$ K the energy monotonically increase with time. At $T=1400$ K it increases up
to $t \approx 80$ ps and then comes to a saturated value. It corresponds to smooth phase transition: the surface melting proliferate
layer by layer, which leads to rapid, but not abrupt changes in the energy.

\begin{figure}

\includegraphics[width=8cm]{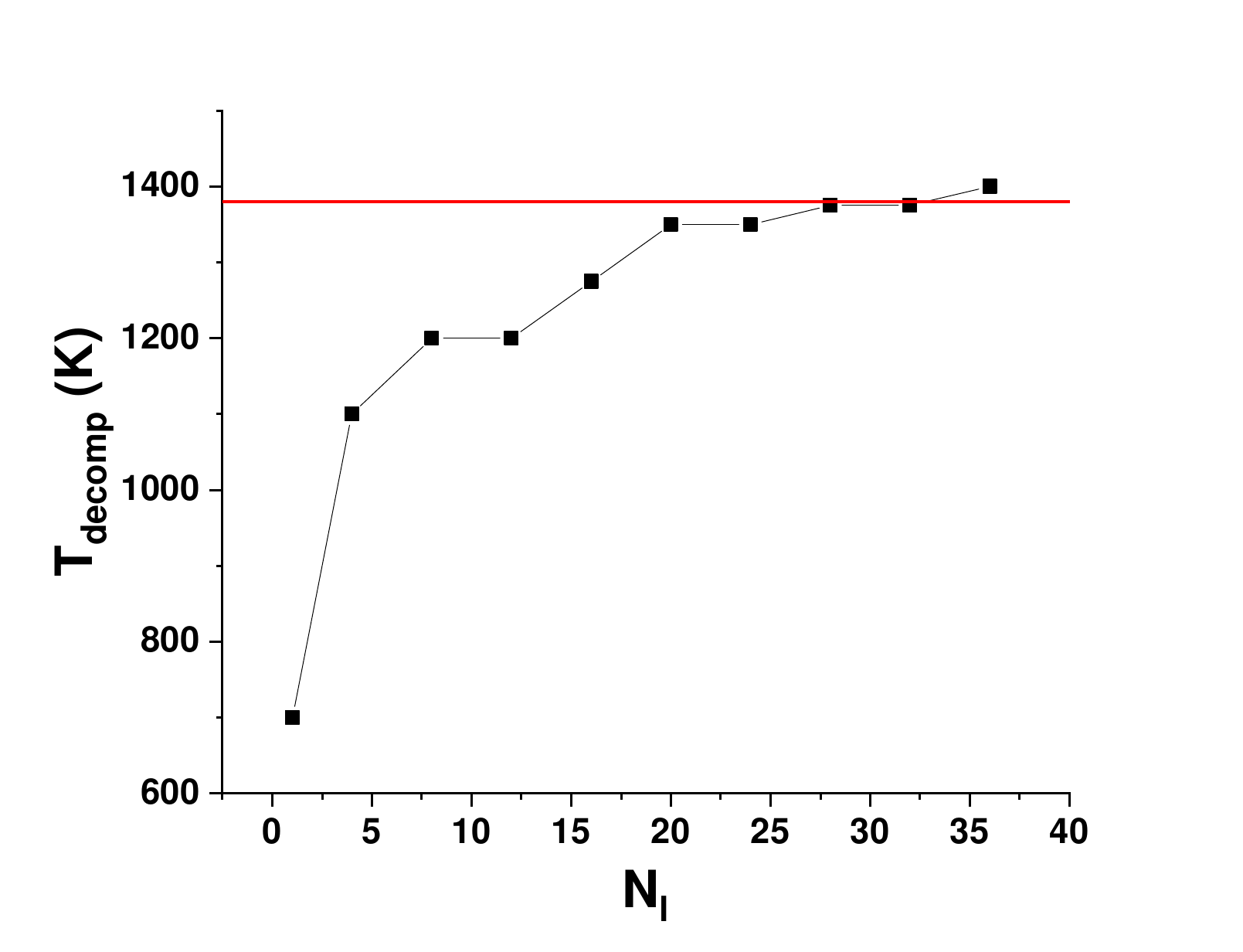}%

\caption{\label{tm} Temperature of thermal decomposition of the crystal structure as a function of the number of silicon layers. Silicene
is denotes as a single layer. The horizontal line shows the melting temperature of a bulk crystal obtained with SNAP model (Ref. \cite{melt-sn}).}
\end{figure}

The temperatures of crystal structure collapse are given in Table I and shown in Fig. \ref{tm}. One can see
that the thermal stability of the films increase with the number of layers. At the film width of
28 layers it comes to the limit of bulk crystal. Further increase of the decomposition temperature most
probably is related to the effects of overheating in simulation.

\subsection{GAP model}

\begin{figure}

\includegraphics[width=8cm]{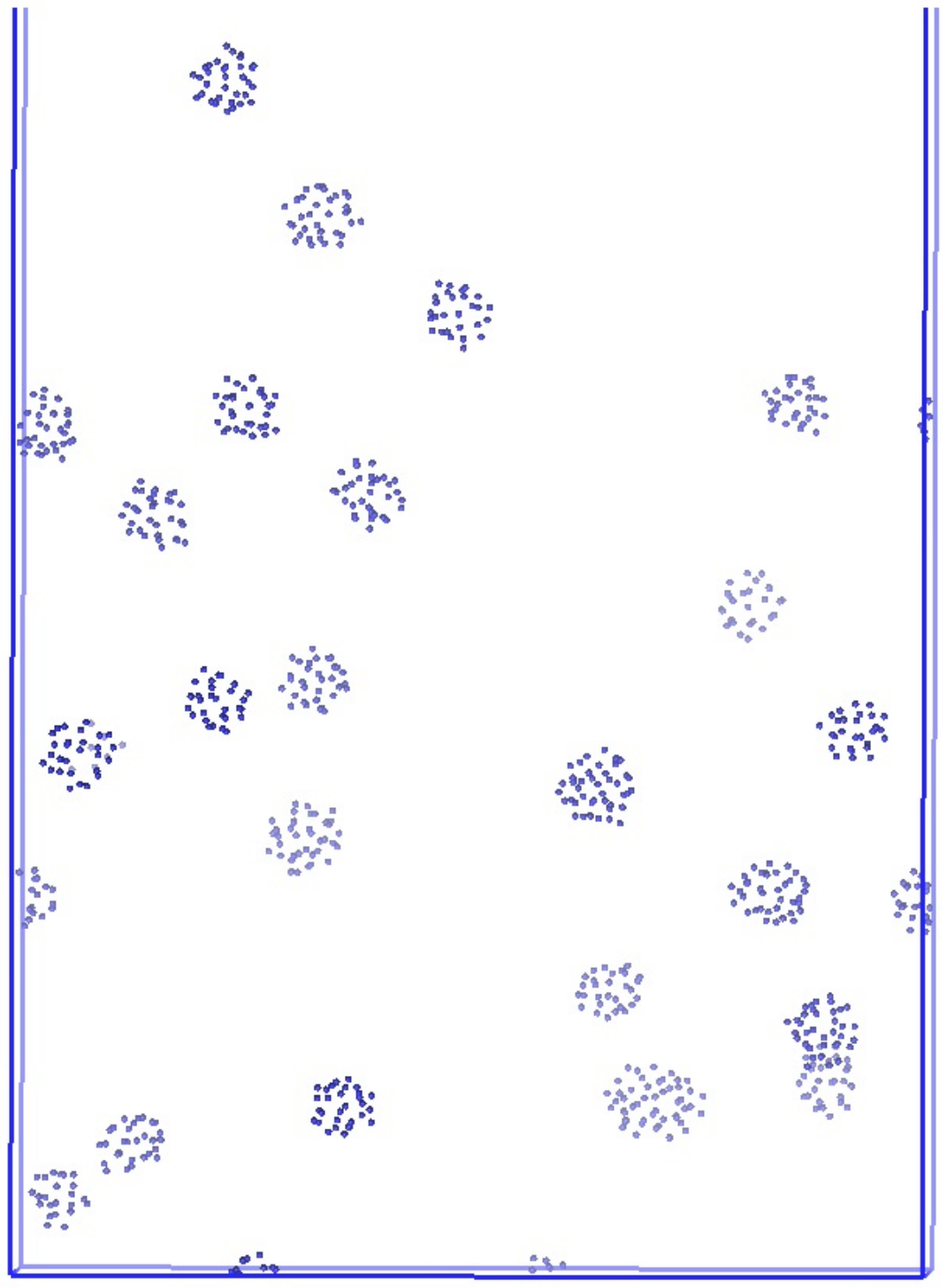}%

\caption{\label{si-gap} A snapshot of the system after collapse of the crystal structure modelled with GAP potential. $T=2000$ K.}
\end{figure}

The simulations of silicene were also performed using the GAP potential. Figure \ref{si-gap} shows a snapshot of the system
after simulation at $T=2000$ K. It can be seen that after the collapse of the silicene structure the atoms gathered into a set of small clusters.
The system came to a state of 'cluster gas': spherical clusters move until a collision with another cluster takes place.
This kind of structure looks to be unphysical. If the GAP potential had shown reasonable behavior for 1800 atoms, then it would have made sense to enlarge the system.

The GAP model was developed to simulate condensed phases of silicon. From these results we conclude that although
GAP potential better than SNAP reproduces the melting line and crystal structures of silicon \cite{melt-sn}, it fails in the
region of gaseous densities. For this reason we do not perform any calculations with this model in the present study.

\section{Conclusions}

In the present work we simulate the thermal stability of silicene and thin silicon films
with two machine-learning potentials: SNAP and GAP. We find that although GAP model is better
in condensed phases of silicon, it fails at low densities and high temperatures. At
the same time the results of SNAP model are in a good agreement with our recent results
for SW empirical model: very thin films decompose at lower temperatures. This reduced stability
can be attributed to the enhanced role of thermal fluctuations in low-dimensional systems,
which more easily disrupt the ordered lattice. The temperature
of decomposition increases with the width of the film. At some value of the width the decomposition
temperature comes to the melting temperature of bulk silicon. In the case of SW model it happens
with the film with 12 layers of silicon, while in the SNAP model 28 layers of silicon are necessary.
Based on the results of these two models we believe that the mechanisms of thermal decomposition
of thin silicon films are reproduced by both models - SW and SNAP - correctly.

This work was carried out using computing resources of the federal
collective usage center "Complex for simulation and data
processing for mega-science facilities" at NRC "Kurchatov
Institute", http://ckp.nrcki.ru, and supercomputers at Joint
Supercomputer Center of the Russian Academy of Sciences (JSCC
RAS). The work on the bulk silicon was supported by Russian Science Foundation (Grant 19-12-00111) and
the work on silicene and silicon films was supported by Russian Science Foundation (Grant  19-12-00092).

\end{document}